\documentclass[osajnl,twocolumn,showpacs,superscriptaddress,10pt]{revtex4-1} %% use 11pt for Applied Optics
\usepackage{amsmath,amssymb,graphicx}
\begin{document}

\title{Nonparaxial shape-preserving Airy beams with Bessel signature}
%\title{On paraxiality of Airy beams with Bessel signature}

\author{Carlos J. Zapata-Rodr\'{\i}guez}
\email{Corresponding author: carlos.zapata@uv.es}
\affiliation{Department of Optics, University of Valencia, Dr. Moliner 50, Burjassot 46100, Spain}

\author{Mahin Naserpour}
\affiliation{Department of Physics, College of Sciences, Shiraz University, Shiraz 71454, Iran}

\begin{abstract}
Spatially accelerating beams that are solutions to the Maxwell equations may propagate along incomplete circular trajectories, after which diffraction broadening takes over and the beams spread out. 
Taking these truncated Bessel wave fields to the paraxial limit, some authors sustained that it is recovered the known Airy beams (AiBs).
Based on the angular spectrum representation of optical fields, we demonstrated that the paraxial approximation rigorously leads to off-axis focused beams %(for instance Gaussian beams) 
instead of finite-energy AiBs.
The latter will arise under the umbrella of a nonparaxial approach following elliptical trajectories in place of parabolas.
Deviations from full-wave simulations appear more severely in beam positioning rather than its local profile.
\end{abstract}


\maketitle %% required

%\section{Introduction}

The accelerating optical beam belongs to a novel class of electromagnetic wave whose peak intensity follows a curved trajectory as it propagates in free space.
Since the first accelerating beam proposed within a paraxial context and propagating along parabolic trajectories \cite{Siviloglou07}, more general classes of solutions have been obtained including elliptical trajectories \cite{Zhang12b,Aleahmad12}, hyperbolic trajectories \cite{Kotlyar14}, and practically any arbitrary trajectory \cite{Froehly11,Mathis13}. 
Irrespective of their acceleration along a convex trajectory, all such beams reportedly display an Airy-shaped intensity across the section transverse to the propagation direction \cite{Greenfield11}.
%A ray optics interpretation of AiBs exists in the literature, allowing to associate the beam trajectory with a longitudinal caustic \cite{Berry79,Vo10}. 

Deriving exact solutions from Maxwell equations is the most general approach, and yields significant theoretical insight. %\cite{Novitsky09,Bandres13c}. 
Recently Kaminer \emph{et al.} presented nonparaxial spatially accelerating shape-preserving beams, which are solutions to the full Maxwell equations and propagate along a circular trajectory reaching angles near $90^\circ$, after which diffraction broadening takes over and the beams spread out \cite{Kaminer12}. 
A key concept is that these beams asymptotically have a high-order Bessel waveform within a given circular sector \cite{Berry69}.
Experimental evidences of these incomplete Bessel beams have been reported in Ref.~\cite{Courvoisier12}.
Since truncation of in-plane Bessel wave fields does not apparently disturb its original waveform in the vicinities of the caustic, one may bring into question the universal association of the Airy function with circular acceleration profiles beyond the principle of stationary phase \cite{Barwick10,Courvoisier12}.

In addition, it is well established that taking these apodized Bessel beams to the paraxial limit recovers the known paraxial AiBs \cite{Kaminer12,Alonso12}.
This statement is primarily based on the fact that a circular trajectory approaches a parabola in the vicinity of its vertex.
We will show that a rigorous application of the paraxial approximation in the plane-wave Fourier expansion of incomplete Bessel wave fields leads to off-axis focused beams instead of finite-energy AiBs.
%Modelling truncation with an appropriate spectral distribution, these paraxial focal waves represent the well-known Gaussian beams.
Furthermore, finite-energy AiBs will arise under the umbrella of a nonparaxial regime by using third-order coefficients, whose trajectories deviate from parabolas.

%\section{In-plane Bessel wavefields}

Let us consider a harmonic wave field propagating in free space, which satisfies the full wave equation $\nabla \times \left(  \nabla \times \mathbf{E} \right) = k^2 \mathbf{E}$,
%\begin{equation}
% \nabla \times \left(  \nabla \times \mathbf{E} \right) = k^2 \mathbf{E} ,
%\end{equation}
where $k = 2 \pi / \lambda$ is the free-space wavenumber.
This equation might be further simplified under the assumption that the wave propagates in the $xy$ plane; therefore waveforms are independent of the spatial coordinate $z$.
Without loss of generality we assume that the polarized wave is transverse electric, set as $\mathbf{E} = \mathbf{z} e_z(x,y)$; validity of our approach is extended to transverse-magnetic polarization by applying the duality theorem.
Finally, the scalar field $e_z$ satisfies the 2D Helmholtz equation, $\left( k^2 + \nabla_t^2 \right) e_z = 0$. 
Solutions to this equation using Bessel functions come out naturally by setting $\nabla_t^2 = \partial_r^2 + r^{-1} \partial_r + r^{-2} \partial_\phi^2$ in a cylindrical coordinate system. 
A general expression for the scalar electric field may be set as 
\begin{equation}
 e_z (r,\phi) = \sum_{m = - \infty}^\infty A_m \psi_m (r,\phi) ,
\label{eq33}
\end{equation}
where $\psi_m = 2 \pi \exp (i m \phi) J_m (k r)$, $J_m (\cdot)$ is a Bessel function of the first kind, and $A_m$ is a complex-valued constant. 
%Equation~(\ref{eq33}) gives a complete solution of the Helmholtz wave equation provided that $e_z$ does not diverge at $r=0$.
Equation~(\ref{eq33}) gives a complete solution provided that $e_z$ does not diverge at $r=0$.

It is convenient to represent the Bessel wave field in the form of a plane-wave Fourier expansion, that is
\begin{equation}
 \psi_m = i^{-m} \int a(\theta) \exp (i m \theta) \exp (i \mathbf{k \cdot r}) \mathrm{d} \theta ,
% \psi_m (r,\phi) = (-i)^m \int a(\theta) \exp (i m \theta) \exp (i \mathbf{k} \mathbf{r}) \mathrm{d} \theta ,
 \label{eq38}
\end{equation}
where the apodization function $a(\theta) = 1$.
Note that a point $\mathbf{r} = \hat{\mathbf{x}} r \cos \phi + \hat{\mathbf{y}} r \sin \phi$ will be set in the Cartesian coordinate system but expressing the direction cosines by means of the radius $r$ (center at $C$) and polar angle $\phi$.
Equation (\ref{eq38}) stands for a isoenergetic superposition of 2D plane waves, modulated azimuthally by a phase-only linear term (and $a$), whose wave vectors $\mathbf{k} = \hat{\mathbf{x}} k \cos \theta + \hat{\mathbf{y}} k \sin \theta$ may be oriented in all directions. 
This field has an intensity profile with a maximum that follows a circular caustic curve of radius $r_m = |m|/k$.

\begin{figure}[htbp]
 \centerline{\includegraphics[width=.9\columnwidth]{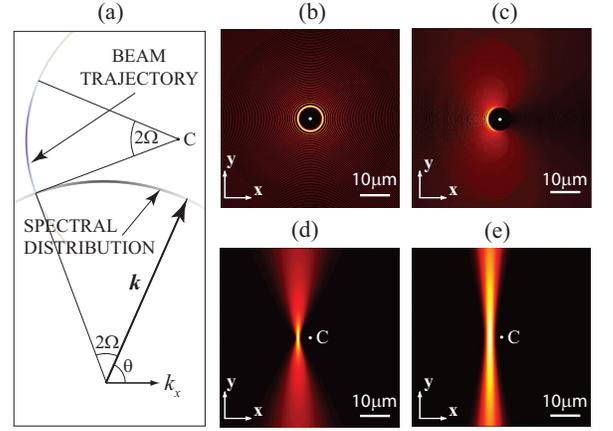}}
 \caption{(a) Spatial spectrum and associated caustic curve for a Gaussian-apodized Bessel field for $l=-50$ and $\Omega = \pi / 10$.
 We also represent the modulus of the field $e_z$ at $\lambda = 500\ \mathrm{nm}$ and different apertures: (b) complete Bessel beam, (c) $\Omega = \pi / 2$, (d) $\Omega = \pi / 10$, and  (e) $\Omega = \pi / 25$.}
 \label{fig01} 
\end{figure}

A relevant feature of the Bessel waveform is in relation with its spatial spectrum.
By means of the Bessel function closure equation \cite{Arfken01}, the spatial spectrum of the wave function $\psi_m$ is retrieved by applying the 2D Fourier transform, namely
\begin{equation}
 F \{ \psi_m \} = 2 \pi i^{-m} k^{-1} \exp (i m \theta) a(\theta) \delta (k - |\mathbf{k}|) ,
\label{eq09}
\end{equation}
%By definition, the 2D-FT of a function $ g(\mathbf{r})$ is derived from $F\{g\}(\mathbf{k}) = \iint g(\mathbf{r}) \exp \left(-i \mathbf{k} \mathbf{r} \right) \mathrm{d}^2 \mathbf{r}$.
where
% the frequency $\mathbf{k} = \hat{\mathbf{x}} k_x + \hat{\mathbf{y}} k_y$ is set in the polar coordinate system as $(|\mathbf{k}|,\theta)$, and 
$\delta$ is the Dirac delta function.
This is also deduced from Eq.~(\ref{eq38}) straightforwardly.
Note that the spectral content of $\psi_m$ is localized along an annulus of radius $|\mathbf{k}| = k$, and again a modulation is driven by $a$ and a phase term, as illustrated in Fig.~\ref{fig01}(a).

%\section{Apertured angular spectra}

Let us now consider an incomplete Bessel wave field of a given order $l$.
For convenience, apodization of the angular spectrum will be governed by a Gaussian distribution.
Therefore we take Eqs.~(\ref{eq33}) and (\ref{eq38}) assuming that $A_m = 0$ for $m \neq l$ and $A_l = 1$ for simplicity.
Moreover the electric field may be set as
\begin{equation}
% e_z = (-i)^l \int a(\theta) \exp \left[i l \theta + i k \left( x \cos \theta + y \sin \theta \right) \right] \mathrm{d} \theta ,
 e_z = \int i^{-l} a(\theta) \exp \left(i l \theta + i k x \cos \theta + i k y \sin \theta \right) \mathrm{d} \theta ,
 \label{eq101}
\end{equation}
being $a(\theta) = (1/{\sqrt{\pi} \Omega}) \exp [-\left({\theta - \pi/2} \right)^2 / \Omega^2 ]$
%\begin{equation}
% a(\theta) = \frac{1}{\sqrt{\pi} \Omega} \exp\left[-\left(\frac{\theta - \pi/2}{\Omega} \right)^2\right] ,
%\end{equation}
the apodizing Gaussian function, having semi-angular aperture $\Omega$ within the domain of integration $|\theta - \pi/2| \le \pi$ and centered at $\theta = \pi / 2$.
%This wave field is also resulting after multiplying the spectral distribution given in Eq.~(\ref{eq09}) at $m = l$ by the apodizing function $a(\theta)$.
As a consequence, phase fronts evolve along a chief direction pointed by the unit vector $\hat{\mathbf{y}}$.
On the other hand, the parametric representation of the associated caustic curve may be set as \cite{Berry69} 
\begin{equation}
 x = (l/k) \sin \theta \ \text{and} \ y = (-l/k) \cos \theta , 
\label{eq108}
\end{equation}
drawing the beam trajectory for angles $\theta$ where $a(\theta)$ takes significant values.
Based on the principle of stationary phase \cite{Papoulis68}, one might estimate a distribution of the field $e_z = a \left( \phi + \sigma_l {\pi}/{2} \right) \psi_l (r,\phi)$,
%\begin{equation}
% e_z (r,\phi) = a \left( \phi + \frac{l}{|l|} \frac{\pi}{2} \right) \psi_l (r,\phi) ,
%\end{equation}
provided that $r \approx r_l$ and $\sigma_l = {l}/{|l|}$.
This represents a sector Bessel wave field centered at $\phi = \pi$ ($\phi = 0$) for $l < 0$ ($l > 0$).

In Fig.~\ref{fig01}(b)--(e) we show the amplitude $|e_z|$ of an incomplete Bessel wave field of order $l=-50$ for different angular apertures, along with the complete Bessel beam.
We observe that the circular caustic curve is also incomplete exhibiting the same angular range $2 \Omega$ of its spectral precursor.
The path coursed by the circular beam has an effective length that is estimated by $2 \Omega r_l = |l| \lambda \Omega / \pi$. %, where $\lambda$ is the wavelength.
Within the paraxial regime, such an arc is necessarily much longer than the wavelength $\lambda$ and simultaneously $\Omega \ll 1$, causing that $|l|$ was limited to extremely high values.
As clearly seen in (e), spatial acceleration ceases to occur for a low-$\Omega$ Gaussian distribution where the beam falls into a rectangular symmetry.
The latter is also reported elsewhere \cite{Chamorro13}.

To get a deeper insight, next it will be assumed that $\Omega \ll 1$.
In this case 
%$\int a(\theta) \mathrm{d} \theta = 1 \pm 0.013$ within the domain of integration $|\theta - \pi/2| \le \pi$; therefore 
it may be accepted to change the limits of integration $|\theta - \pi/2| \le \pi$ in Eq.~(\ref{eq101}).
Such a diffraction integral is now conveniently established on an unbounded domain, a fact that is ambiguously interrelated with paraxiality of wave propagation along the $y$ axis.
As an illustration, let us consider the expansions 
\begin{subequations}
\begin{eqnarray}
 \cos \theta & \approx & - \theta' + \theta'^3 / 6 , \label{eq100a} \\
 \sin \theta & \approx & 1 - \theta'^2 / 2 , \label{eq100b}
\end{eqnarray}
\label{eq100}
\end{subequations}
being $\theta' = \theta - \pi/2$.
Finite Taylor series~(\ref{eq100}) are straightforwardly applicable to the wave vector $\mathbf{k}$, and are accurate about $\theta' = 0$ with an error term $O(\theta'^4)$.
Now it is clear that paraxial wave propagation requires a sufficiently low angular aperture $\Omega$ enabling accurate use of Eqs.~(\ref{eq100}) with only quadratic polynomials.

%\subsection{Paraxial regime}

\begin{figure}[htbp]
 \centerline{\includegraphics[width=.9\columnwidth]{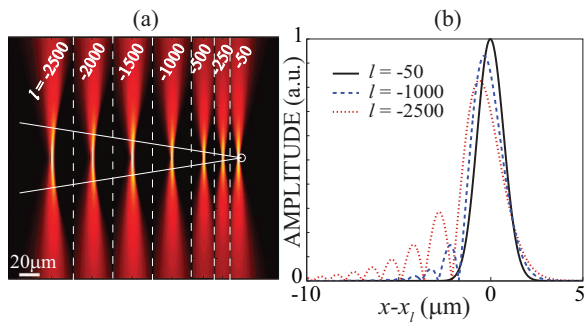}}
 \caption{(a) Field $|e_z|$ at $\lambda = 500\ \mathrm{nm}$ and evaluated from Eq.~(\ref{eq101}) for $\Omega = \pi / 20$ at different Bessel orders $l$.
 The solid white lines crossing at origin determine the boundaries of a sector of angle $2 \Omega$ where caustics arise in all such wave fields.
 The circular caustic signature of the Bessel wave field for $l=-50$ is also represented in white.
 (b) Distribution of the field $|e_z|$ at $y = 0$ and at different orders.} 
 \label{fig02} 
\end{figure}

Following our course of reasoning, the dispersion relation derived from Eq.~(\ref{eq09}), $k_x^2 + k_y^2 = k^2$ where $\mathbf{k} = \hat{\mathbf{x}} k_x + \hat{\mathbf{y}} k_y$, will be reduced to a quadratic function in the paraxial regime, namely $k_y = k - {k_x^2}/{2 k} + O[k_x]^4$.
Since the shape and angular aperture of the spatial spectrum seems to coincide with the trajectory of the beam, differing by a rotation of $90^\circ$ (and scaled appropriately), % as seen in Fig.~\ref{fig01}(a), 
one may conceive that the caustic curve governed by application of a quadratic approach in Eqs.~(\ref{eq100}) will also take the form of a parabola.
The latter occurs if it is implemented in the parametric equation (\ref{eq108}) giving % provided that $O(\theta'^3)$, namely
\begin{equation}
 k^2 y^2 + 2 k l x = 2 l^2 .
\label{eq109}
\end{equation}
However, a farseeing analysis carries us to a different result.
Applying an error function $O(\theta'^3)$ in the angular spectrum representation of the field given in Eq.~(\ref{eq101}), the caustic reduces to a single point, $x = x_l = l/k$ and $y = 0$ \cite{Berry80}.
%Beyond this approximation up to $O(\theta'^4)$ it finally yields $x = (x_l / 2) + (r_l / 2) \cos 2 \alpha$ and $y = (1/\sqrt{2}) r_l \sin 2 \alpha$, being $\theta' = \sqrt{2} \tan \alpha$.
Beyond this approximation up to $O(\theta'^4)$ the caustic follows the curve 
\begin{equation}
 (x,y) = \left( \frac{x_l}{2} + \frac{x_l}{2} \cos 2 \alpha , \frac{x_l}{\sqrt{2}} \sin 2 \alpha \right) ,
\label{eq107}
\end{equation}
being $\tan \alpha = \theta' / \sqrt{2}$.
Equation~(\ref{eq107}) represents an ellipse whose center is set at $(x,y) = (x_l/2,0)$ and whose semi-major axis and semi-minor axis are given by $r_l/2$ and $r_l/\sqrt{2}$, respectively.

When we neglect terms in Eqs.~(\ref{eq100}) of order higher than the 2nd degree, the diffraction integral (\ref{eq101}) has an analytical solution.
This yields
\begin{equation}
 e_z (x,y) = \frac{\exp(i k y) }{\chi(y)} \exp \left[-\frac{(l - k x)^2 \Omega ^2}{4 \chi^2(y)}\right] ,
 \label{eq102}
\end{equation}
where $\chi(y) = \sqrt{1 + {i k y \Omega^2}/{2}}$.
%\begin{equation}
% \chi(y) = \sqrt{1 + \frac{i k y \Omega^2}{2}} .
%\end{equation}
Note that Eq.~(\ref{eq102}) represents the electric field of a paraxial 2D Gaussian beam whose waist is set at $y = 0$.
Irrespective of the order $l$, the Gaussian spot size at this focal plane is $w_G = 2/\Omega k$, and consequently its Rayleigh range is given by $y_R = 2 / \Omega^2 k$.
Moreover, the Gaussian beam is shifted laterally at $x = x_l$.
Under this approach, the role of $\exp(i l \theta)$ in Eq.~(\ref{eq101}) is functionally identical to the phase term produced by a blazed grating achieving a maximum diffraction efficiency at $(x_l,0)$.
% – also called echelette grating (from French échelle = ladder) – is a special type of diffraction grating. It is optimized to achieve maximum grating efficiency in a given diffraction order. For this purpose, maximum optical power is concentrated in the desired diffraction order 
Such a point belongs to the circular caustic of the Bessel signature, and therefore the \emph{undeviating} intensity peak of the Gaussian beam is tangent to such a caustic.
This is illustrated in Fig.~\ref{fig01}(e) and also in Fig.~\ref{fig02}(a) and (b) for $l = - 50$.

We have seen above that diffraction governs wave localization in low-$\Omega$ low-$l$ incomplete Bessel fields.
In order to estimate the angular aperture for validity of the \emph{quadratic} approximation, we will confront the propagation distance $2 y_R$ (confocal parameter) of the Gaussian beam, governed by diffraction, and the length $2 \Omega r_l$ of the trajectory that may be inferred from optical-ray basis.
Under these circumstances, the quadratic regime clearly applies provided that $\Omega r_l \ll y_R$, 
%as illustrated in Fig.~\ref{fig02}(a), 
that is if $\Omega \ll \Omega_l$ being $\Omega_l = (2 / |l|)^{1/3}$.

Figure~\ref{fig02}(a) and (b) show the electric field corresponding to incomplete Bessel wave fields of semi-aperture $\Omega = \pi / 20$ at different orders.
Note that $\Omega_l = \Omega$ is approximately satisfied if $l = - 516$.
At $l = -50$ the wave profile looks alike the electric field of the Gaussian beam given in Eq.~(\ref{eq102}).
On the other hand, optical beams clearly deviate from the Gaussian profile at higher values of $|l|$ where spatial acceleration takes place.

%\subsection{Weakly-nonparaxial effects}

%Next we consider a higher value of $\Omega$ (approaching $l^{-1/3}$) characterizing the apodizing function $a(\theta)$.
%This fact makes the apertured beam accelerates, establishing a higher-order approximation for the wave vector $\mathbf{k}$. 
%In this case, we will take the series given in Eq.~(\ref{eq100}) terminating after the 3rd order Taylor polynomial, that is $\cos \theta$ will be now substituted by $- \theta' + \theta'^3 / 6$.
Next we establish a higher-angle approximation for the wave vector $\mathbf{k}$ in the spectral representation of $e_z$.
%This fact makes the apertured beam accelerates, establishing a higher-order approximation for the wave vector $\mathbf{k}$. 
For this purpose, we will take the series in Eq.~(\ref{eq100}) terminating after the 3rd order Taylor polynomial, that is $\cos \theta$ will now include the term $\theta'^3 / 6$.
Inclusively under these conditions, the plane-wave Fourier expansion (\ref{eq101}) may be set using a simple analytical expression, namely
\begin{equation}
 e_z = \frac{2 \sqrt{\pi}}{\Omega} M(x) \exp \left\{ i \left[ k y + \alpha(x,y) \right] \right\} \mathrm{Ai} \left[ \zeta(x,y) \right] ,
 \label{eq104}
\end{equation}
where $\mathrm{Ai}(\cdot)$ is the Airy function, and
\begin{subequations}
\begin{align}
 M(x)           &= \exp \left({i 2 \pi}/{3} \right) \left( {2}/{k x} \right)^{1/3} , \\
 \gamma(x,y)    &= \left( {k y}/{2} - i \Omega^{-2} \right) M^2(x) , \\
 \zeta(x,y)     &= \left( l - k x \right) M(x) - \gamma^2(x,y) , \\
 \alpha(x,y)    &= \zeta (x,y) \gamma (x,y) + \gamma^3 (x,y) / 3 .
\end{align}
\label{eq105}
\end{subequations}
We point out that Eq.~(\ref{eq104}) represents the field of a finite-energy AiB out of the paraxial regime.
Since the main peak of the Airy function is roughly centered at the origin, wave localization is estimated at points satisfying $\zeta(x,y) = 0$ (in the limit $\Omega \to \infty$).
This yields the equation of an ellipse, $y^2 = -2 x (x - x_l)$, coinciding with Eq.~(\ref{eq107}).
Contrarily to what is commonly assumed, incomplete Bessel wave fields drive finite-energy AiBs with elliptical trajectories instead of parabolas.
Note that this acceleration behavior refers to the trajectories of the local beam intensity features in spite of the fact that the center of gravity of these waves remains invariant in accord with Ehrenfest's theorem \cite{Siviloglou07}.

%\section{Paraxial waveforms}

For large Bessel orders, it may happens that $\Omega_l$ approached (or even fell behind) $\Omega$ and concurrently the latter takes values that might be commonly considered valid in the paraxial regime.
In this case one expect that the field $e_z = \exp(i k y) \Psi(x,y)$ related to the AiB would have a spatial distribution in a way that $\Psi$ satisfied the paraxial wave equation, $( \partial_x^2 + 2 i k \partial_y ) \Psi = 0$. 
It is simply retrieved if we substitute $M(x)$ by $M(x_l)$, which matches $\Omega_l$ except for a phase-only factor,
%valid in the region of interest, 
while other terms in Eq.~(\ref{eq104}) and (\ref{eq105}) remain the same.
The electric field $e_z$ set in this form is in agreement with the solutions to the paraxial wave equation originally reported in Ref.~\cite{Siviloglou07}.
Furthermore, omitting terms of higher order than the 2nd degree in $\Omega^{-1}$ we found $\zeta = \left( 2/l \right)^{1/3} \left( l - k x  - k^2 y^2 / 2 l \right)$.
The equation $\zeta(x,y) = 0$ for the caustic curve now leads to a parabola with points satisfying $y^2 = -2 x_l (x - x_l)$, which naturally coincides with Eq.~(\ref{eq109}).
%In addition,
%\begin{equation}
% \alpha(x,y) = \frac{k y ( l- k x)}{l} - \frac{2 i (l - k x)}{l \Omega ^2} + \frac{4 i k^2 y^2}{l^2 \Omega ^2} - \frac{2 k^3 y^3}{3 l^2} + O(\Omega ^{-4}) .
%\end{equation}

At this point we are in condition to confirm the significance of the angle $\Omega_l$.
Taking Eq.~(\ref{eq104}) and (\ref{eq105}) within the paraxial approach given above, the main peak size of the AiB is approximately $w_{Ai} = 2 / |M| k$.
%, provided that $\Omega_l < \Omega$.
Keeping out this wavelet from the whole diffracted waveform, it gives shape to a (Fourier-limited) wave packet of far-field beam angle \cite{Siegman86} $2 / k w_{Ai} \equiv \Omega_l$.
In other words, the uttermost spot of the AiB is effectively generated by a part of the spatial spectrum having an angular aperture $\Omega_l$. 
Thus sidelobes of the Airy profile are attributed to interference effects from remaining component of the AiB spatial spectrum \cite{Zapata11}.
If $\Omega < \Omega_l$, however, the angular spectrum is completely devoted to the formation of a focused beam of size $w_G = 2 / \Omega k$.
%This explanatory remark may be straightforwardly applied to more general accelerating beams and nonparaxial regimes.

\begin{figure}[htbp]
 \centerline{\includegraphics[width=.85\columnwidth]{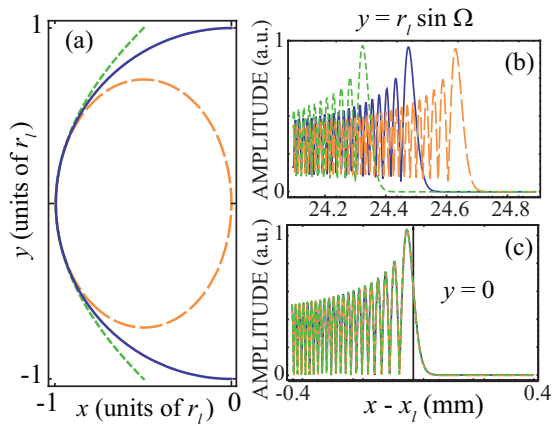}}
 \caption{(a) Caustic curves for the Bessel signature (solid line), the parabolic AiB (dotted line) and the nonparaxial AiB (dashed line).
 Field $|e_z|$ in the plane (b) $y = r_l \sin \Omega$ and (c) $y = 0$ at $\lambda = 500\ \mathrm{nm}$ for an incomplete Bessel field of $l = -25 \times 10^6$ and $\Omega = \pi / 20$.} 
% \caption{Solutions with very high index $l$ to see differences between weakly-nonparaxial effects and its paraxial waveform.} 
 \label{fig03} 
\end{figure}

We point out that for extremely large orders $l$, the existence of AiBs may span a wide domain all along the spatial coordinate $x$, even considering sufficiently low angles $\Omega \ll 1$ (however $\Omega_l \ll \Omega$).
This occurs from the merging point (vertex) of the curvilinear caustics given in Eqs.~(\ref{eq108}), (\ref{eq109}) and (\ref{eq107}) to their edges where they move far away, seen in Fig.~\ref{fig03}(a). 
Near the edges, the field characterizing these (either paraxial or nonparaxial) AiBs will deviate significantly from full-wave estimations.
For illustration, in Fig.~\ref{fig03}(b) we show the AiB pattern at one of the edges, $y = r_l \sin \Omega$, assuming the paraxial approach and the 3rd-order nonparaxial regime, together with the numerical simulation obtained from Eq.~(\ref{eq101}) [full-wave analysis disregarding approximations (\ref{eq100})].
It is clear that deviations are more severe in beam positioning rather than their local profiles, at least close to the main peak.
This is consistent with the fact that at the vertex plane all analytical and numerical estimations essentially exhibit the same field [see Fig.~\ref{fig03}(c)].
Finally, getting along the circular caustic curve will lead to further departures of the Airy patterns (not shown in Fig.~\ref{fig03}).

We conclude that the accelerating performance for an AiB with Bessel signature and its degree of paraxiality is limited not only by the aperture $\Omega$ of its spatial spectrum but also its characteristic width $\Omega_l$ governing wave concentration in the vicinity of the caustic curve.
%We may conclude that the paraxial regime for an accelerating beam with Bessel signature is limited not only by the aperture $\Omega$ of its spatial spectrum but also its local width $\Omega_l$ governing wave concentration in the vicinity of the caustic curve.
An analogous remark may be straightforwardly applied to more general accelerating beams.

%\section*{Acknowledgments}    

This research was funded by the Spanish Ministry of Economy and Competitiveness under the project TEC2011-29120-C05-01.

%\bibliography{Bibliography}

%

%\bibliographystyle{osajnl}
%\bibliography{Bibliography}

\end{document}